\documentclass[aps,pre,twocolumn,showpacs,groupedaddress,superscriptaddress, amsmath,amssymb]{revtex4-1}

\usepackage[pdftex]{graphicx}

\usepackage{amsmath}
\usepackage{amssymb}
\usepackage{amsfonts}
\usepackage{dcolumn}% Align table columns on decimal point
\usepackage{bm}% bold math
\usepackage{color}

\begin{document}

\title{Onset of two collective excitations in the transverse dynamics of a simple fluid}

\author{Eleonora Guarini}
%\thanks{Corresponding Author.}
%\email{guarini@fi.infn.it}
\affiliation{Dipartimento di Fisica e Astronomia, Universit\`a degli Studi di Firenze, via G. Sansone 1, I-50019 Sesto Fiorentino, Italy}

\author{Martin Neumann}
\affiliation{Fakult\"{a}t f\"{u}r Physik der Universit\"{a}t Wien, Kolingasse 14-16, A-1090 Wien, Austria}

\author{Alessio De Francesco}
\affiliation{CNR-IOM \& INSIDE@ILL c/o Operative Group in Grenoble (OGG) F-38042 Grenoble, France
and Institut Laue Langevin (ILL), F-38042 Grenoble, France}

\author{Ferdinando Formisano}
\affiliation{CNR-IOM \& INSIDE@ILL c/o Operative Group in Grenoble (OGG) F-38042 Grenoble, France
and Institut Laue Langevin (ILL), F-38042 Grenoble, France}

\author{Alessandro Cunsolo}
\affiliation{Department of Physics, University of Wisconsin at Madison, 1150 University Avenue, Madison, 53706 Wisconsin, USA}

\author{Wouter Montfrooij}
\affiliation{Department of Physics and Astronomy, University of Missouri, Columbia, 65211 Missouri, USA}

\author{Daniele Colognesi}
\affiliation{Consiglio Nazionale delle Ricerche, Istituto di Fisica Applicata ``Nello Carrara'', via Madonna del Piano 10, I-50019 Sesto Fiorentino, Italy}

\author{Ubaldo Bafile}
\affiliation{Consiglio Nazionale delle Ricerche, Istituto di Fisica Applicata ``Nello Carrara'', via Madonna del Piano 10, I-50019 Sesto Fiorentino, Italy}

\begin{abstract}

A thorough analysis of the transverse current autocorrelation function obtained by molecular dynamics simulations of a dense Lennard-Jones fluid reveals that even such a simple system is characterized by a varied dynamical behavior with changing length scale. By using the exponential expansion theory, we provide a full account of the time correlation at wavevectors $Q$ between the upper boundary of the hydrodynamic region and $Q_{\rm p}/2$, with $Q_{\rm p}$ the position of the main peak of the static structure factor. In the $Q$ range studied we identify and accurately locate the wavevector at which shear wave propagation starts to take place, and show clearly how this phenomenon may be represented by a damped harmonic oscillator changing, in a continuous way, from an overdamped to an underdamped condition. The decomposition into exponential modes allows one to convincingly establish not only the crossover related to the onset of transverse waves but, surprisingly, also the existence of a second pair of modes equivalent to another oscillator that undergoes, at higher $Q$ values, a similarly smooth over- to underdamped transition.

\end{abstract}

%\date{\today}

\maketitle

In the field of the microscopic dynamics of fluids, increasing attention has been paid in the last two decades to the study of shear waves. The propagation of both longitudinal and transverse acoustic waves depends not only on the thermodynamic state of the fluid but also on the length scale (i.e., on the wavevector modulus $Q=|{\bf Q}|$) at which these processes are probed. However, contrary to what happens for ordinary sound, where propagating longitudinal waves also exist in the $Q \to 0$ limit, even in very dilute gases, these dependencies have a peculiar character in the transverse case because the propagation of shear waves in fluids takes place only when both the density and $Q$ exceed certain threshold values \cite{levesque1973}.

The existence of such a transition between the absence and presence of propagating transverse excitations has attracted much interest due to its evident link with the viscoelastic nature of liquids \cite{levesque1973, balucani, cunsolo}, whereby the system responds in a solidlike way to a perturbation of sufficiently high frequencies and wavevectors, while slower and longer-wavelength perturbations dissipate through viscous relaxation processes. The connection of the threshold $Q$-value (usually referred to in the literature as $Q_{\rm gap}$) with the relaxation time that Maxwell first introduced to account for viscoelasticity has been even recently discussed \cite{yang2017}.

One of the dynamical quantities that are also sensitive to transverse particle motions (i.e., orthogonal to ${\bf Q}$) is the velocity autocorrelation function \cite{bellissima2017, guarini2017}. Moreover, a widely debated issue concerns the visibility of transverse excitations in experimental determinations \cite{hosokawa2009, giordano2010, hosokawa2013, zanatta2015, guarini2020} and simulations \cite{marques2015, delrio2017, delrio2017a} of the spectrum of density fluctuations, i.e., of the dynamic structure factor $S(Q,\omega)$, in liquid metals. However, the quantities that directly capture the essence of the transverse dynamics are the transverse-current time autocorrelation function $C_{\rm T}(Q,t)$ and its frequency spectrum $\tilde{C}_{\rm T}(Q,\omega)$ \cite{nota1}. Therefore, investigating the conditions that allow for the propagation of shear excitations requires the analysis of transverse-current data at various $Q$ values. However, before concentrating on specific features such as the excitation frequencies and damping rates and their $Q$ dependence, and on the determination of $Q_{\rm gap}$, a valid modeling of $C_{\rm T}(Q,t)$ or $\tilde{C}_{\rm T}(Q,\omega)$ should, in first place, provide an accurate representation of available data in the entire time or frequency ranges. Here data means simulation results, since $\tilde{C}_{\rm T}(Q,\omega)$ is currently not accessible by spectroscopic techniques able to probe the picosecond and nanometer scales.

In this Letter, we pursue this goal using molecular dynamics (MD) simulation data for a dense  Lennard-Jones (LJ) fluid at the (slightly supercritical) temperature $T = 1.35$ and number density $n = 0.8$. A few details about the simulations can be found in the Supplemental Material (SM) \cite{suppl}. Throughout this paper we will use the standard dimensionless variables defined via the LJ parameters $\epsilon$ and $\sigma$ and the particle mass $m$, but for ease of notation we will omit to mark reduced variables with the usual asterisk \cite{reduced}. $C_{\rm T}(Q,t)$, computed from the simulated dynamics of $N=24805$ particles, was analyzed in the $Q$ range between 0.2 and 3.4 in steps of 0.2. This range covers the first half of the region delimited by the position $Q_{\rm p}$ of the main peak of the static structure factor $S(Q)$, located in this system at $Q_{\rm p}=6.76$.

For the analysis of the MD data we apply the exponential expansion theory (EET) \cite{barocchi2012, barocchi2013, barocchi2014} which allows for excellent descriptions of various correlation functions and spectra of interest for the self \cite{bellissima2017, guarini2017} and collective dynamics \cite{guarini2020, guarini2021}. EET states that any autocorrelation function can  be expressed as a series of exponential terms (called modes). Thus, for $t\ge 0$, we write, at each $Q$ value 

\begin{equation}\label{eq:EET}
C_{\rm T}(Q,t)=C_{\rm T}(Q,0)\sum_{j=1}^{\infty}I_j\exp(z_jt),
\end{equation}

\noindent where both $I_j$ and $z_j$ can either be real or complex, with ${\rm Re}\,z_j<0$. A real mode describes the exponential decay of a relaxation process, while a pair of complex conjugate modes accounts for a propagating excitation with damping coefficient $-{\rm Re}\,z_j$ and frequency $|{\rm Im}\,z_j|$. In both cases, we shall refer to $-{\rm Re}\,z_j$ as the ``damping'' of the mode. In Eq.\ (\ref{eq:EET}), $I_j$ and $z_j$ are dependent on $Q$, although this will not be explicitly indicated in the following. We refer the reader to Refs.\ \cite{bellissima2017, guarini2017, guarini2020} for details on the application of EET, where a fitting procedure is applied to determine the parameters $z_j$ and $I_j$ of a small number $p$ of modes to which the sum in Eq.\ (\ref{eq:EET}) effectively reduces. Here we only note that $p-1$ constraints have been imposed to the amplitudes $I_j$ in order to enforce the correct short time behavior of the fitted $C_{\rm T}(Q,t)$ \cite{guarini2020}, including $\sum_{j=1}^{p}I_j=1$.

Hydrodynamic theory predicts that in the $Q\to 0$ limit $\tilde{C}_{\rm T}(Q,\omega)$ has a Lorentzian shape with a half width at half maximum (HWHM) given by $(\eta/n)Q^2$ \cite{absolut}, where $\eta$ is the shear viscosity \cite{balucani}. Such a Lorentzian spectrum obviously corresponds to retaining only one term in Eq.\ (\ref{eq:EET}), when $C_{\rm T}(Q,t)$ decays through a purely diffusive process. The hydrodynamic behavior should be obtained as the long wavelength limit of the more complex dynamics observed at higher $Q$.

\begin{figure}
\resizebox{0.45\textwidth}{!}
{\includegraphics[viewport=2.5cm 3cm 23cm 18.5cm]{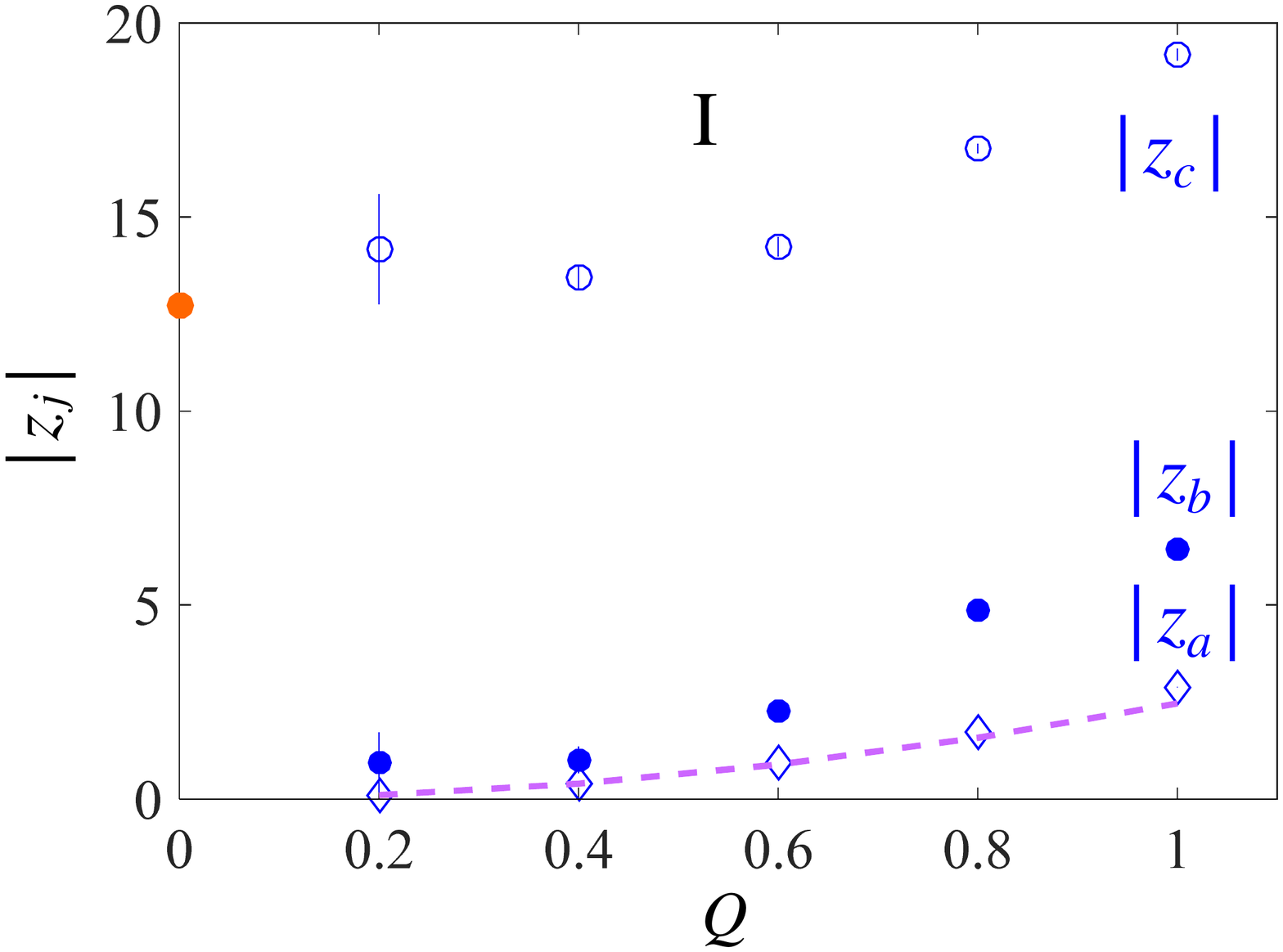}}
\caption{Damping coefficients $|z_a|$, $|z_b|$, and $|z_c|$ of the three modes fitted to $C_{\rm T}(Q,t)$ in range I. The dashed curve displays the $(\eta/n)Q^2$ hydrodynamic behavior. The orange dot at $Q=0$ indicates the value of $1/\tau_{\rm M}$.}
\label{fig:fig1} 
\end{figure}

\begin{figure}
\resizebox{0.45\textwidth}{!}
{\includegraphics[viewport=2.5cm 3.5cm 17cm 28.5cm]{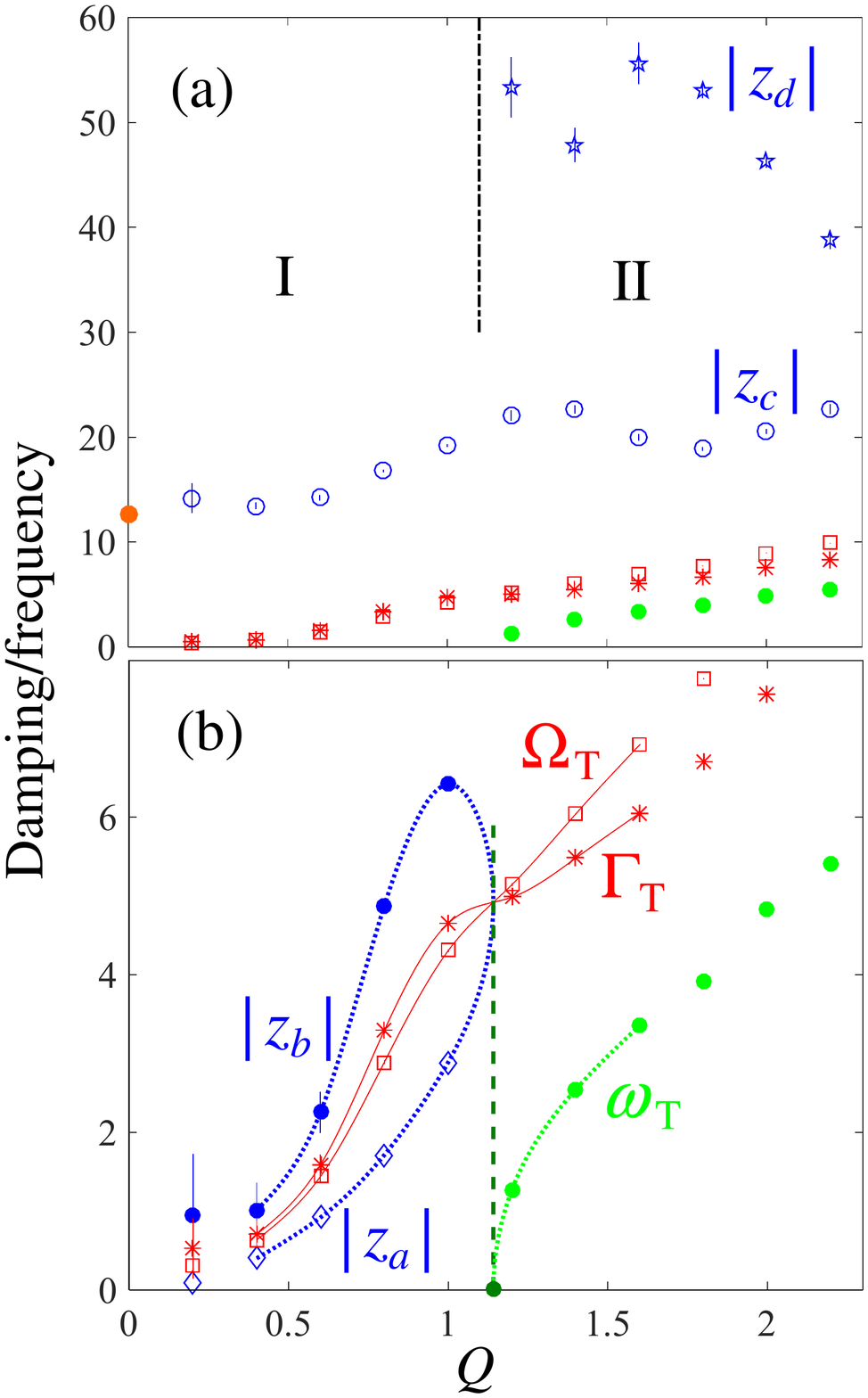}}
\caption{Damping and frequency parameters from the fits to $C_{\rm T}(Q,t)$ in ranges I and II. (a) The dampings of real modes $c$ and $d$ in range II are shown as blue circles and blue stars, respectively. $\Omega_{\rm T}$ (red squares) and $\Gamma_{\rm T}$ (red asterisks), as calculated from the dampings $z_a$ and $z_b$ in range I and directly fitted in range II (see text), are shown. The dispersion curve $\omega_{\rm T}$ is displayed as green dots. The orange dot is the same as in Fig.\ \ref{fig:fig1}. (b) Zoom of the low frequency region of frame (a), showing also $|z_a|$ and $|z_b|$ from Fig.\ \ref{fig:fig1}. The red lines are spline interpolations of $\Omega_{\rm T}(Q)$ and $\Gamma_{\rm T}(Q)$ across the transition from the overdamped to the underdamped regime. The green dashed line marks the value of $Q_{\rm gap}$. The dotted blue lines show the behavior of $|z_a|$ and $|z_b|$ in range I near the transition point. The dotted green line shows the beginning of the dispersion curve $\omega_{\rm T}(Q)$.}
\label{fig:fig2} 
\end{figure}

\begin{figure}
\resizebox{0.45\textwidth}{!}
{\includegraphics[viewport=2.5cm 8cm 17cm 26cm]{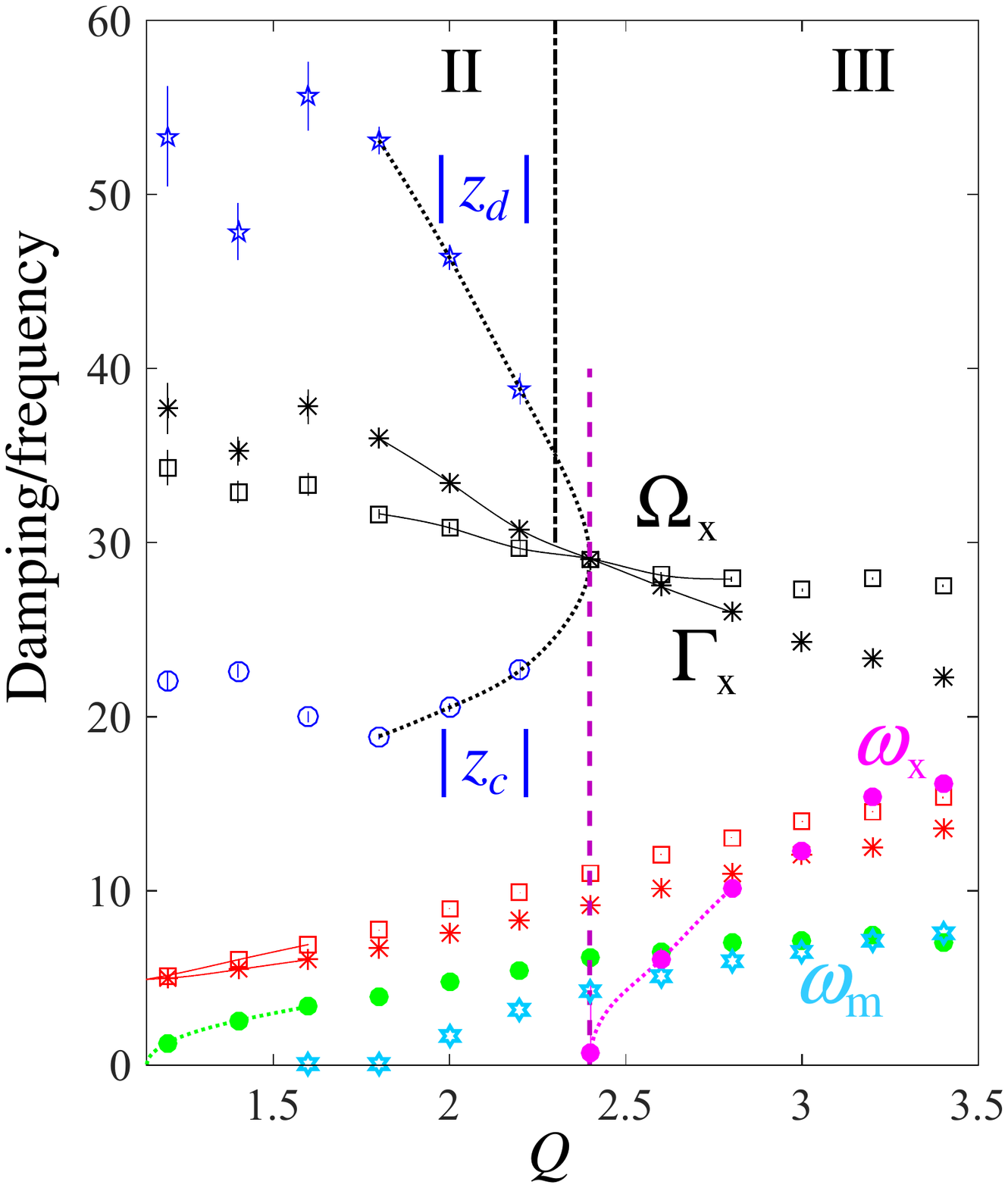}}
\caption{The damping and frequency parameters from the fits to $C_{\rm T}(Q,t)$ in ranges II and III. Besides the curves already displayed in Fig.\ \ref{fig:fig2}, here we show $\Omega_{\rm x}$ (black squares) and $\Gamma_{\rm x}$ (black asterisks), as calculated from the dampings $z_c$ and $z_d$ in range II and directly fitted in range III (see text). The dispersion curve $\omega_{\rm x}(Q)$ of the second excitation is displayed with magenta full circles. The black lines are spline interpolations of $\Omega_{\rm x}(Q)$ and $\Gamma_{\rm x}(Q)$ across the transition from the overdamped to the underdamped regime of the second excitation. The purple dashed line marks the $Q$ at which the transition takes place. The dotted black lines show the behavior of $|z_c|$ and $|z_d|$ in range II near the transition point. The dotted magenta line shows the initial part of the second dispersion curve. The cyan hexagons indicate the $Q$-dependence of the position $\omega_{\rm m}$ of the maximum in each $\tilde{C}_{\rm T}(Q,\omega)$ spectrum: the first two points with $\omega_{\rm m}=0$ are reported in order to indicate that the maximum in the spectrum starts having a nonzero position in frequency somewhere between $Q=1.8$ and $Q=2.0$.}
\label{fig:fig3} 
\end{figure}

\begin{figure}
\resizebox{0.45\textwidth}{!}
{\includegraphics[viewport=2.5cm 2cm 23cm 21cm]{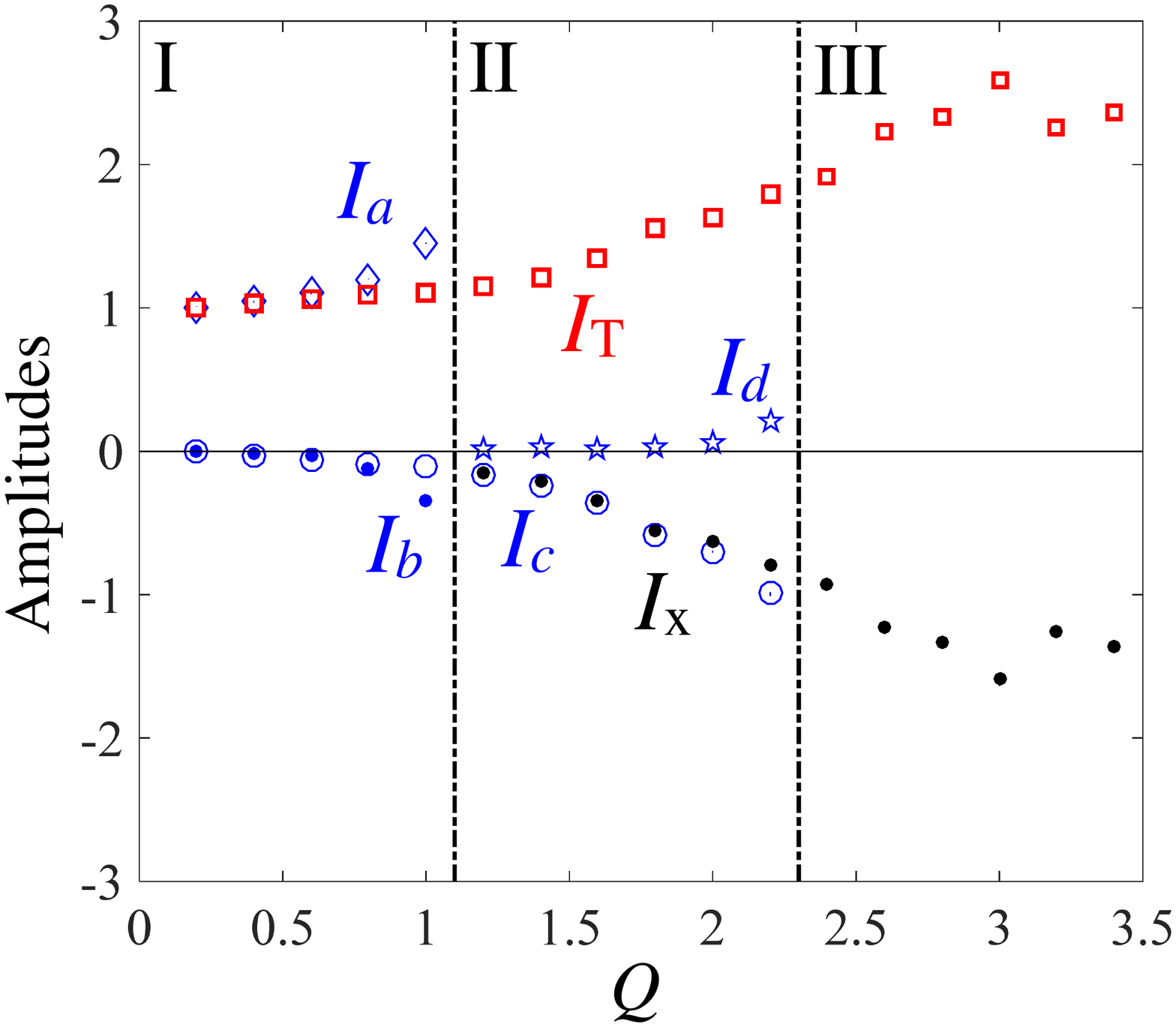}}
\caption{$Q$ dependence of the amplitudes of the fitted modes. Blue symbols refer to the real mode amplitudes: $I_a$ (diamonds) and $I_b$ (dots) in range I, $I_c$ (circles) in ranges I and II, $I_d$ (stars) in range II. The amplitudes $I_{\rm T}$ of the transverse oscillator and $I_{\rm x}$ of the second oscillator are also shown as red squares and black dots, respectively.}
\label{fig:fig4} 
\end{figure}

The range of wavevectors included in this study is divided in three parts, labeled as I, II, and III in the following, where different sets of exponential modes are required to accurately fit the time dependence of $C_{\rm T}(Q,t)$. Specifically, we find that in a rather narrow $Q$ interval a variety of dynamical behaviors occurs, smoothly transitioning from one $Q$-regime into the other. The best-fitting parameters $z_j$ are reported in Figs. \ref{fig:fig1}-\ref{fig:fig3}, and their respective amplitudes $I_j$ are displayed in Fig.\ \ref{fig:fig4}. (Examples of the fits are shown in Figs. S1-S3 of the SM.)

In range I ($0.2\le Q \le 1.0$), very good fits to $C_{\rm T}(Q,t)$ are obtained when three real exponential terms are included in Eq.\ (\ref{eq:EET}). The three modes, identified by subscripts $a$, $b$ and $c$, decay to zero with damping constants $|z_a|$, $|z_b|$ and $|z_c|$, shown in Fig.\ \ref{fig:fig1}, which are the HWHMs of the three Lorentzian lines composing the spectrum. Mode $a$, with the smallest damping, is the dominant one ($I_a \ge 1$ in Fig.\ \ref{fig:fig4}), while $I_b$ and $I_c$ are negative and vanish for $Q\to 0$ but increase slowly, in absolute value, with growing $Q$. We find that $|z_a|$ grows initially as $Q^2$ with a prefactor in full agreement with the quantity $(\eta/n)Q^2$ mentioned above (the LJ viscosity is taken from Ref.\ \cite{meier2002}). We also remark that, while $|z_b|$ starts from zero too, the width $|z_c|$ of the third Lorentzian line is only weakly dependent on $Q$ and does not vanish for $Q\to 0$, where it approaches a value close to the reciprocal of the Maxwell time $\tau_{\rm M}$ \cite{nota2}. Overall, the low-$Q$ dynamics complies with the hydrodynamic limit, but the detailed analysis in terms of three exponential modes elucidates a slight yet clearly progressive deviation from the limiting behavior.

In range II ($1.2\le Q \le 2.2$), the above three-mode model becomes insufficient for an accurate description of the data, and four modes are now required, two of which are complex and two real. One of the latter is labeled $c$ because both its damping $|z_c|$ and amplitude $I_c$ evolve very smoothly from the corresponding parameters of the $c$ mode of range I, indicating that the same relaxation process is actually present in both $Q$ ranges (see Figs.\ \ref{fig:fig2}(a) and \ref{fig:fig4}) . The other, labeled as $d$, has a very large damping $|z_d|$ and, despite its almost negligible amplitude, is necessary to reach a high fit quality (see Fig.\ S2 of the SM).

The remaining two modes fitted to $C_{\rm T}(Q,t)$ in range II form a pair of complex conjugate terms in Eq.\ (\ref{eq:EET}). It is well known  \cite{bafile2006, montfrooij2021} that such a pair describes the dynamics of a damped harmonic oscillator \cite{nota3} characterized by an undamped frequency $\Omega_{\rm T}$, a damping $\Gamma_{\rm T}$, and, when $\Omega_{\rm T}>\Gamma_{\rm T}$, an actual oscillation frequency $\omega_{\rm T}=\sqrt{\Omega_{\rm T} ^2-\Gamma_{\rm T}^2}$. For the two modes, again denoted as $a$ and $b$ for the reasons to be explained below, one has $z_{a,b}=-\Gamma_{\rm T}\pm i\omega_{\rm T}$. Therefore, the two complex modes contribute to the total $C_{\rm T}(Q,t)$ with an oscillatory underdamped component corresponding to a propagating collective excitation.

However, if the above condition were reversed, with $\Omega_{\rm T}<\Gamma_{\rm T}$, overdamping would occur and no oscillation would appear. The two modes would become real, with damping constants given by $z_{a,b}=-\Gamma_{\rm T}\pm \sqrt{\Gamma_{\rm T}^2-\Omega_{\rm T}^2}$. In both damping conditions it is seen that $\Omega_{\rm T}^2=z_{a}z_{b}$ and $\Gamma_{\rm T}=-(z_{a}+z_{b})/2$. It is then natural to check whether modes $a$ and $b$ determined in range I can be interpreted as representing an overdamped oscillator which evolves smoothly into the underdamped one defined in range II by $\Omega_{\rm T}$ and $\Gamma_{\rm T}$. Therefore, also in range I, we define $\Omega_{\rm T}^2=z_a z_b$ and $\Gamma_{\rm T}=-(z_a+z_b)/2$ and observe that both $\Omega_{\rm T}(Q)$ and $\Gamma_{\rm T}(Q)$, reported in Fig.\ \ref{fig:fig2}, have a smooth $Q$ dependence in the whole range $0.2 \le Q \le 2.2$. Remarkably, the sum of the amplitudes of the two modes $I_{\rm T}=I_a+I_b$ also displays a very smooth crossing of the boundary between range I and II (see Fig.\ \ref{fig:fig4}).
   
We can thus confidently recognize modes $a$ and $b$ as present in both ranges I and II. Since mode $a$ was seen to account for the transverse dynamics in the hydrodynamic limit, we conclude that the onset of the propagating transverse excitation is properly described as the transition of an oscillator from a low-$Q$ overdamped condition ($\Omega_{\rm T}<\Gamma_{\rm T}$) to an oscillating one ($\Omega_{\rm T}>\Gamma_{\rm T}$). This justifies the use of a single suffix ``T'' to label the transverse modes. As shown in Fig.\ \ref{fig:fig2}b, by means of a spline interpolation of $\Omega_{\rm T}(Q)$ and $\Gamma_{\rm T}(Q)$, the point where they cross each other, which by definition is $Q_{\rm gap}$, is determined with great accuracy to be $Q_{\rm gap}=1.14$, and the dispersion curve of the transverse excitation

\begin{equation}\label{eq:omegaT}
\omega_{\rm T}(Q)=\sqrt{\Omega_{\rm T} ^2(Q)-\Gamma_{\rm T}^2(Q)}
\end{equation}

\noindent is found, where we explicitly indicated the $Q$ dependence of {\it both} quantities under the square root. From the interpolated values of $\Omega_{\rm T}$ and $\Gamma_{\rm T}$ in range I the behavior of the real-mode dampings $|z_a|$ and $|z_b|$ is obtained in the vicinity of the transition point, where they approach each other and eventually merge into $\Gamma_{\rm T}$.  

Finally, in range III ($2.4\le Q \le 3.4$), best fits are also obtained with a four-exponential model for $C_{\rm T}(Q,t)$, but now all modes are complex, two of which provide an impressively smooth continuation to larger $Q$ of the propagating transverse excitation described by $\Omega_{\rm T}(Q)$, $\Gamma_{\rm T}(Q)$ and $I_{\rm T}(Q)$ (see Figs.\ \ref{fig:fig3} and \ref{fig:fig4}). The other complex pair of modes is seen to arise from the modes denoted as $c$ and $d$ in range II, with exactly the same kind of transition process from over- to underdamping discussed before for the transverse excitation. Thus, the data display the onset of a second collective excitation characterized by the undamped frequency $\Omega_{\rm x}(Q)$ and the damping $\Gamma_{\rm x}(Q)$, which in range II are obtained as $\Omega_{\rm x}^2= z_c z_d$ and $\Gamma_{\rm x}=-(z_c+z_d)/2$ and in range III are determined directly as fit parameters. Therefore, at the transition point, located at $Q$=2.4 as the crossing point of spline interpolations of $\Omega_{\rm x}$ and $\Gamma_{\rm x}$, a second dispersion curve $\omega_{\rm x}(Q)=\sqrt{\Omega_{\rm x} ^2(Q)-\Gamma_{\rm x}^2(Q)}$ emerges, with a larger slope than that of $\omega_{\rm T}$. 

The transverse dispersion curve $\omega_{\rm T}(Q)$ cannot be determined unless it is obtained from the frequencies of the oscillatory components of $C_{\rm T}(Q,t)$ at each $Q$ value. This requires the exponential-mode description, either in the form used here or in the so-called generalized collective modes approach \cite{mryglod1998}; however, in the latter case, the modes are not fitted to the data. By contrast, the dispersion curve is usually obtained as the frequency position $\omega_{\rm m}(Q)$ of the maxima of the individual $\tilde{C}_{\rm T}(Q,\omega)$ spectra. The result of such a procedure is also shown in Fig.\ \ref{fig:fig3} (cyan hexagons). Since in the present case the spectra show a double-peaked top for $Q\ge 2.0$ only, this method gives for $Q_{\rm gap}$ a value between $1.8$ and $2.0$, quite larger than the already found $Q_{\rm gap}=1.14$. The values of $\omega_{\rm m}(Q)$ not only fail to reproduce the correct frequencies $\omega_{\rm T}$, but also do not allow to detect the second excitation $\omega_{\rm x}$ in range III. In fact, for $Q \ge 2.0$ the shape of $\tilde{C}_{\rm T}(Q,\omega)$ does not show any visible difference in passing from the one-excitation to the two-excitation dynamics (see Figs.\ S2 and S3 of the SM).

As far as the second excitation is concerned, a double structure in $\tilde{C}_{\rm T}(Q,\omega)$ has been found in {\it ab initio} simulations of some liquid metals \cite{delrio2017b}, though in some cases only at very high pressures \cite{bryk2015a, jakse2019}. Interestingly, in Ref.\ \cite{guarini2017} it has also been shown that the transverse current spectra of liquid gold display a clear shoulder at frequencies rather close to the maximum frequency of the sound dispersion curve. Moreover, a mixing of longitudinal and transverse excitations of $C_{\rm T}(Q,\omega)$ was observed in water \cite{sampoli97} and methanol \cite{bellissima2016}. Here, it is useful to note that, at our highest $Q$, the dispersion curve $\omega_{\rm x}(Q)$ attains values that already exceed the transverse frequency $\omega_{\rm T}$ by a factor $\approx 2$. Therefore, although we cannot make any ultimate statement or claim on the nature of the second excitation, we suggest that it may be related to the longitudinal dynamics even in this much simpler fluid.

In conclusion, we have shown that an analysis of $C_{\rm T}(Q,t)$ in terms of fitted exponential modes reveals a rich dynamical behavior and enables a consistent interpretation through the simple concept of damped harmonic oscillators, discussed in detail in Refs.\ \cite{bafile2006, montfrooij2021}, whose intuitive meaning is that relaxation and propagation phenomena are driven by the competition between ``elastic'' forces and viscous dissipation, represented by the undamped frequencies ($\Omega_{\rm T}$ or $\Omega_{\rm x}$) and the damping coefficients ($\Gamma_{\rm T}$ or $\Gamma_{\rm x}$), respectively. In this way, detailed properties of the transverse collective dynamics are revealed, including the appearance of a second excitation so far undetected in simple nonmetallic fluids, besides the expected emergence of the transverse mode propagation. Moreover, a very accurate determination of the threshold $Q$ values is made possible. This work shows that in the $Q$ evolution of $C_{\rm T}(Q,t)$, the whole intensity, initially associated with a single hydrodynamic decay channel, smoothly redistributes among four modes which give rise, in pairs, to the onset of two propagating waves. An essential requirement for such an analysis is the accurate description of the entire time dependence of $C_{\rm T}(Q,t)$ in terms of exponential modes, where the choice of the number and nature of the modes to be fitted must be made at each $Q$ by duly comparing the fit quality of different models, while avoiding unjustified overparametrizations.

\begin{acknowledgments}
This research was funded by Ministero dell'Istruzione dell'Universit\`a e della Ricerca Italiano (Grant No. PRIN2017-2017Z55KCW). We are greatly indebted to Walter Penits for technical support of the simulations. 
\end{acknowledgments}


\begin{thebibliography}{}
\bibitem{nota1} For the definition of $C_{\rm T}(Q,t)$ we follow Ref.\ \cite{balucani}.
\bibitem{reduced} In reduced units, temperature and density are given by $k_{\rm B}T/\epsilon$ and $n \sigma^3$, respectively, while wavevector and angular frequency are expressed by $Q \sigma$ and $\omega \sqrt{m \sigma^2/\epsilon}$.
\bibitem{absolut} In absolute units, the halfwidth is given by $\eta/(nm) Q^2$ \cite{balucani}.
\end{thebibliography}
\end{document}

\title{Supplemental Material\\
for\\
Onset of two collective excitations in the transverse dynamics of a simple fluid }

\author{Eleonora Guarini}
\affiliation{Dipartimento di Fisica e Astronomia, Universit\`a degli Studi di Firenze, via G. Sansone 1, I-50019 Sesto Fiorentino, Italy}

\author{Martin Neumann}
\affiliation{Fakult\"{a}t f\"{u}r Physik der Universit\"{a}t Wien, Kolingasse 14-16, A-1090 Wien, Austria}

\author{Alessio De Francesco}
\affiliation{CNR-IOM \& INSIDE@ILL c/o Operative Group in Grenoble (OGG) F-38042 Grenoble, France
and Institut Laue Langevin (ILL), F-38042 Grenoble, France}

\author{Ferdinando Formisano}
\affiliation{CNR-IOM \& INSIDE@ILL c/o Operative Group in Grenoble (OGG) F-38042 Grenoble, France
and Institut Laue Langevin (ILL), F-38042 Grenoble, France}

\author{Alessandro Cunsolo}
\affiliation{Department of Physics, University of Wisconsin at Madison, 1150 University Avenue, Madison, 53706 Wisconsin, USA}

\author{Wouter Montfrooij}
\affiliation{Department of Physics and Astronomy, University of Missouri, Columbia, 65211 Missouri, USA}

\author{Daniele Colognesi}
\affiliation{Consiglio Nazionale delle Ricerche, Istituto di Fisica Applicata ``Nello Carrara'', via Madonna del Piano 10, I-50019 Sesto Fiorentino, Italy}

\author{Ubaldo Bafile}
\affiliation{Consiglio Nazionale delle Ricerche, Istituto di Fisica Applicata ``Nello Carrara'', via Madonna del Piano 10, I-50019 Sesto Fiorentino, Italy}

\maketitle

The MD simulations were performed with the leapfrog algorithm in the isokinetic ensemble, using a cubic simulation volume linked-list cells, a force cutoff of $r_c=6.5$, and a timestep of $\delta t =$ 0.002 \cite{allen}. The correlation functions reported are averages over 10 independent sub-runs of $10^7$ timesteps each, where averages over the sub-runs were used to estimate the error bars on the net correlation. The number density was $n=0.8$, and the number of particles $N=24805$ was chosen such that the minimum wavenumber compatible with the simulation cell has the value $Q_{\rm{min}}=0.2$, and only wavevectors parallel to the three Cartesian axes were considered.

The fits were performed in the time domain. It is known \cite{allen,hansen} that the use of periodic boundary conditions in MD may produce spurious effects in the calculated correlations beyond the so-called recurrence time $t_{\rm R}=(N/n)^{1/3}/c_{\rm s}$, i.e. the time required by a density wave to propagate over the box length at the adiabatic sound speed $c_{\rm s}$. With $c_{\rm s}=6.22$ \cite{meier2002} one has $t_{\rm R}=5$, which defines the maximum time value to be used for the fits. However, with the exception of the lowest $Q$'s, $C_{\rm T}(Q,t)$ substantially decays to zero in a time shorter than $t_{\rm R}$ and the time range for the fit is accordingly reduced to avoid inclusion in the fit of noisy and meaningless data displaying only statistical fluctuations around zero.

\begin{figure*}
\resizebox{0.99\textwidth}{!}
{\includegraphics[trim=2cm 11cm 2cm 2cm]{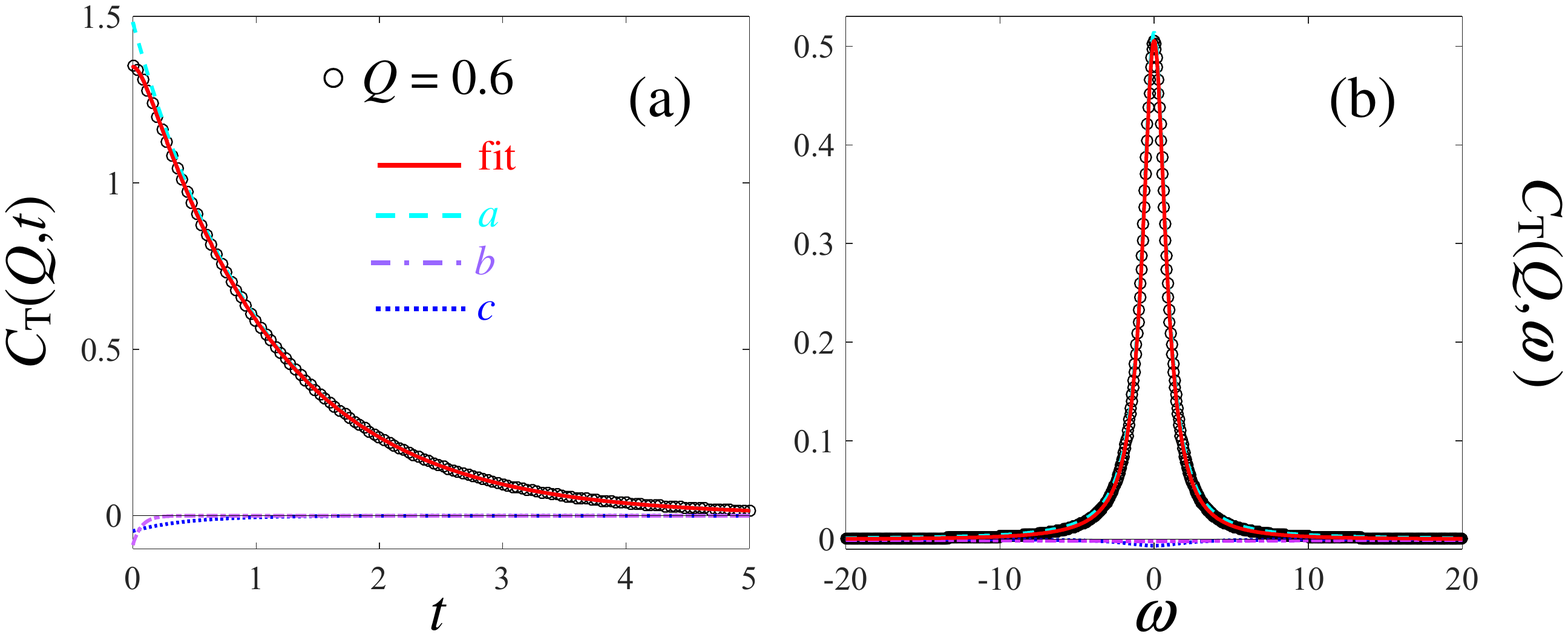}}
\caption{(a) Transverse current autocorrelation at a $Q$ representative of region I (black circles). The fit result and its components are specified in the legend according to the labeling of the (real) modes present in this wavevector region and described in the main text. (b) Corresponding spectrum and fit components. }  
\label{zonaI}
\end{figure*}

\begin{figure*}
\resizebox{0.70\textwidth}{!}
{\includegraphics[trim=2cm 2.5cm 2cm 2cm]{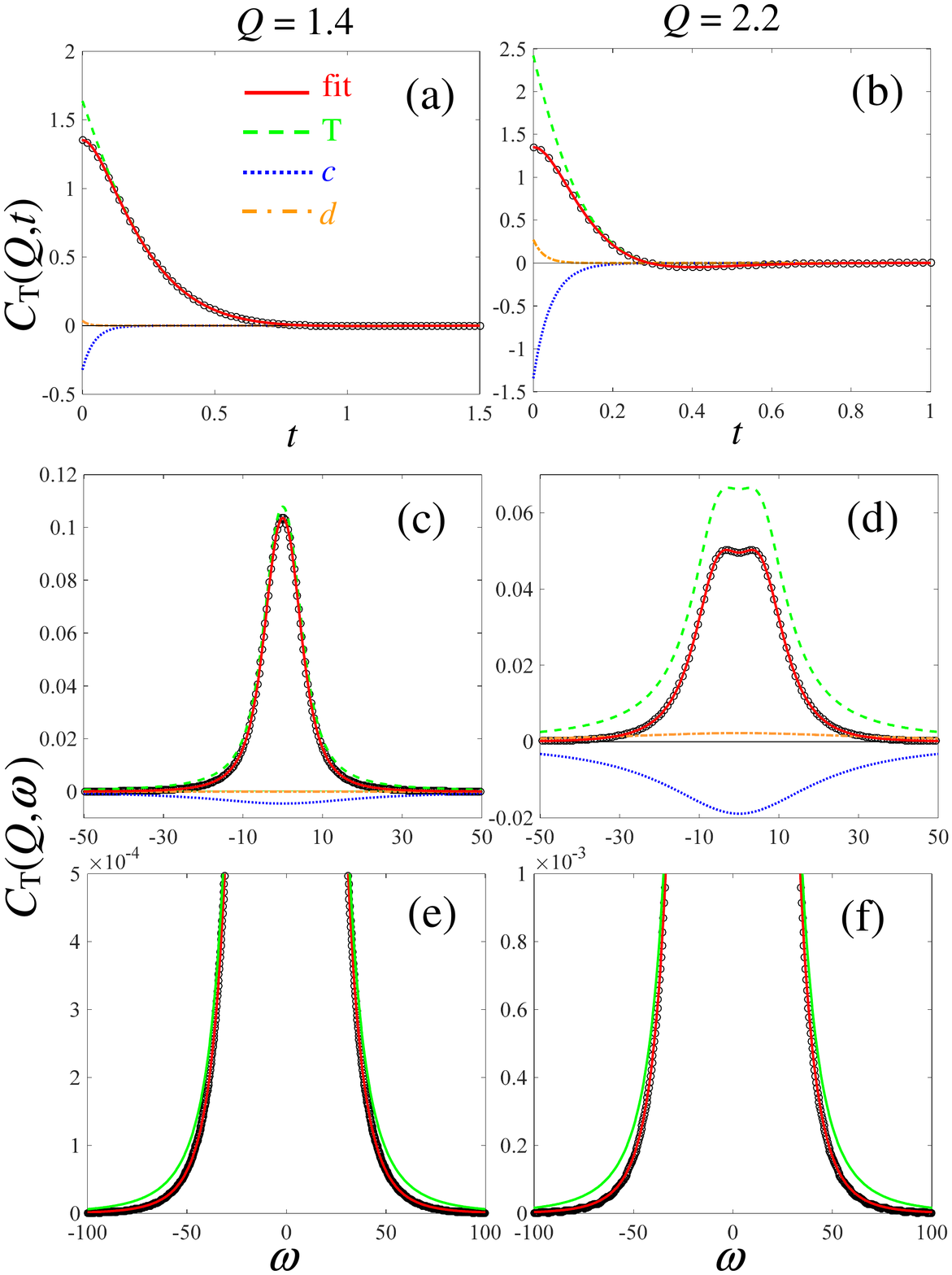}}
\caption{(a)-(d) $C_{\rm T}(Q,t)$ and its spectrum at two $Q$ values in region II. Here, the overdamped oscillator of region I (composed of the modes $a$ and $b$ in Fig.\ \ref{zonaI}) has become underdamped, and transverse waves propagate in the fluid (dashed green T component). One real mode of region I continues to be present (dotted blue curves), together with another real mode (dot-dashed orange curves), and they represent the components of a second oscillator which is in overdamped conditions in region II. (e)-(f) Comparison of the fits shown in panels (c) and (d) (red curve passing through the data points) with fits performed by excluding mode $d$ (green curve). }  
\label{zonaII}
\end{figure*}

\begin{figure*}
\resizebox{0.99\textwidth}{!}
{\includegraphics[trim=2cm 11cm 2cm 2cm]{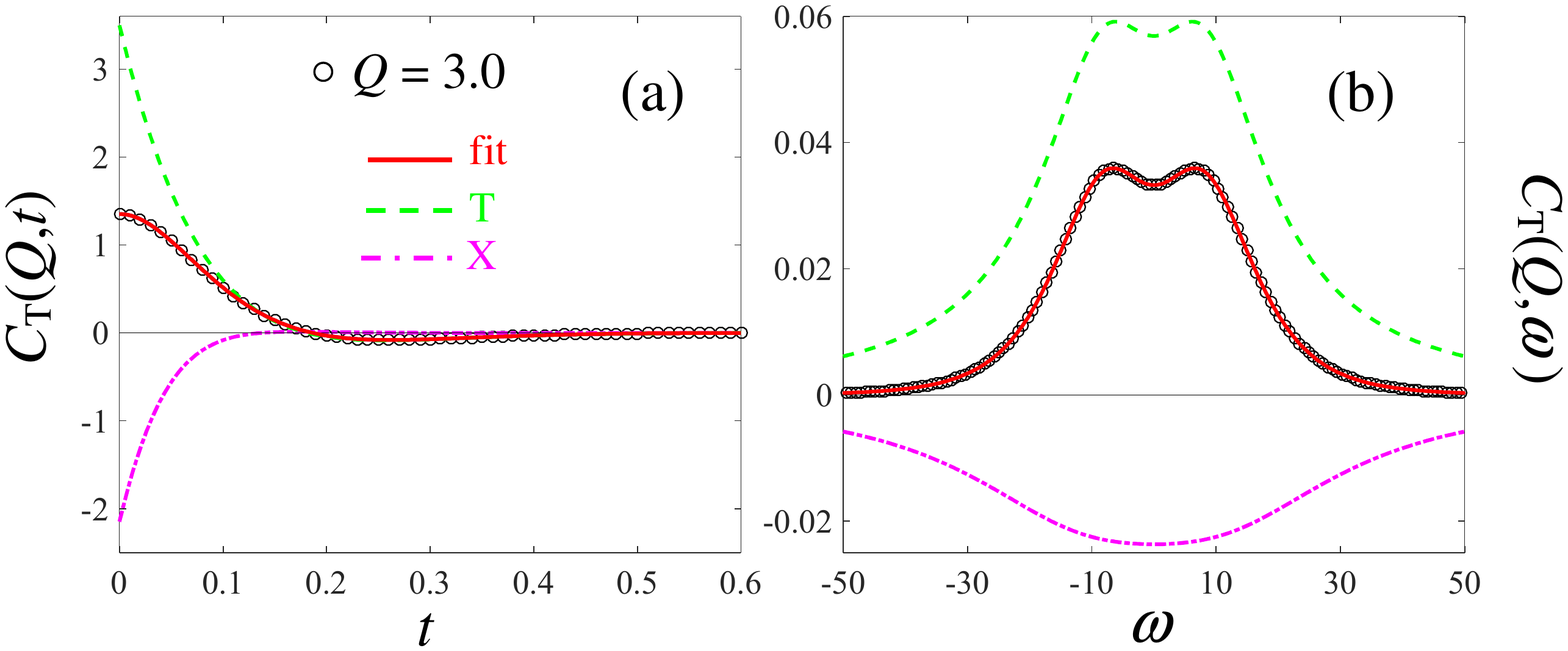}}
\caption{Same as Fig.\ \ref{zonaI}, but for region III, where both oscillators are in underdamped conditions. The transverse propagating waves of region II are still present (dashed green), but a second excitation is clearly seen to emerge in the fluid (magenta curves) in this wavevector region. }  
\label{zonaIII}
\end{figure*}

The MD data for  $C_{\rm T}(Q,t)$, the fitted exponential-mode model, and its various components are shown in three figures, one for each of the three $Q$ regions introduced in the main paper. Each figure also shows the corresponding spectrum and its components. Figure \ref{zonaI} refers to $Q=0.6$ (in range I), and Fig.\ \ref{zonaIII} refers to $Q=3.0$ (in range III). For range II the results pertaining to two wavevector values ($Q=1.4$ and $2.2$) are presented in Fig.\ \ref{zonaII} in order to show how a double peak develops in the spectrum, and also to notice that at certain $Q$ values, despite the clear oscillatory component found by the fit procedure, the spectrum does not (apparently) display any inelastic feature. Moreover, the last two panels of Fig.\ \ref{zonaII} show how a fit model excluding the low-intensity mode $d$ of range II fails in accounting for the tails of the spectra. In particular, with the simpler (green) model with one real mode only, at $Q$=1.4 and 2.2 the reduced $\chi^2$ is, respectively, 5 and 200 times larger than that of the (red) four-mode fits.  \\

\newpage

\newpage
